\documentclass{article}
\usepackage{amsmath}
\usepackage{amsfonts}
\usepackage{amssymb}

\usepackage[pdftex]{graphicx} 
\usepackage{asymptote}

\usepackage[hmargin=2.5cm,vmargin=3.0cm]{geometry}
\usepackage[font=small,labelfont=bf, margin=1cm]{caption}

\renewcommand{\thefootnote}{\fnsymbol{footnote}}  

\begin{document}
\setlength{\textheight}{8.0truein}    

\thispagestyle{empty}
\setcounter{page}{1}


\vspace*{0.88truein}



\centerline{\bf
REDUCTION FROM NON-INJECTIVE HIDDEN}
\vspace*{0.035truein}
\centerline{\bf  SHIFT PROBLEM TO INJECTIVE HIDDEN SHIFT PROBLEM}
\vspace*{0.37truein}
\centerline{\footnotesize
MIRMOJTABA GHARIBI}
\vspace*{0.015truein}
\centerline{\footnotesize\it Cheriton School of Computer Science and Institute for Quantum Computing}
\baselineskip=10pt
\centerline{\footnotesize\it University of Waterloo, 200 University Avenue West}
\baselineskip=10pt
\centerline{\footnotesize\it Waterloo, Ontario N2L 3G1,
Canada}
\vspace*{10pt}

\vspace*{0.21truein}

\abstract{
We introduce a simple tool that can be used to reduce non-injective instances of the hidden shift problem over arbitrary group to injective instances over the same group. In particular, we show that the average-case non-injective hidden shift problem admit this reduction. We show similar results for (non-injective) hidden shift problem for bent functions. We generalize the notion of influence and show how it relates to applicability of this tool for doing reductions. In particular, these results can be used to simplify the main results by Gavinsky, Roetteler, and Roland about the hidden shift problem for the Boolean-valued functions and bent functions, and also to generalize their results to non-Boolean domains (thereby answering an open question that they pose).
}{}{}

\vspace*{10pt}

\vspace*{3pt}

\section{Introduction} 
\noindent       
After Shor's discovery of an efficient quantum algorithm for factoring and the discrete log problems, research on the hidden subgroup problem (HSP) attracted many scholars in the field  \cite{shor97}. HSP is a framework which includes factoring and the discrete log in itself  \cite{childs10}. Despite the early success in finding a solution for the abelian HSP, achieving a similar result has proven to be hard for the non-abelian case  \cite{childs10}. HSP is important since solutions for it over the dihedral group and the symmetric group will yield solutions to some lattice problems and graph isomorphism respectively  \cite{beals97,ettinger99,hoyer97,boneh95}. In both cases, we have a non-abelian instance of HSP.

The hidden shift problem (also known as the hidden translation problem) was defined in the works of   \cite{vandam03,friedl03}. Interesting problems can be stated as a hidden shift problem, most notably this includes hidden subgroup problem over dihedral group, which is equivalent to the hidden shift problem over $\mathbb{Z}_{N}$, and graph isomorphism, which can be cast as a hidden shift problem over $S_n$  \cite{ettinger00,kuperberg05,childs07}. The study of the hidden shift problem can give an arguably more natural view to tackle the graph isomorphism problem  \cite{childs07}. 

In the injective hidden shift problem, we are given two injective functions over some group $G$ that are simply a shifted version of each other. The task is to output such a shift. More formally, let $f, g : G \rightarrow S$ be two injective functions such that, for some unique $s\in G$, it holds that 
\begin{equation}
f(x)=g(sx) \text{ for all } x\in G\,. \label{eq1} 
\end{equation}
The goal is to find the hidden shift $s$.

Relaxing the requirement for the functions to be injective, will lead to a variant of the problem. We call this new problem, the non-injective hidden shift problem. We restrict the problem to the instances with non-periodic functions, so that the hidden shift will be unique.

By lower bounds on the query complexity of the unstructured search problem, a worst case solution to the non-injective hidden shift problem cannot be obtained  \cite{bennett97}. Imposing restrictions on the instances makes the non-injective hidden shift problem more tractable. In particular, in this paper, we are concerned with the average case non-injective hidden shift problem and also the hidden shift problem for bent functions.

The non-injective hidden shift problem has been studied for a variety of functions. Efficient quantum algorithm for solving the hidden shift problem when $f:\mathbb{Z}_{p}\rightarrow \{-1,0,1\}$ is the  Legendre symbol is presented in the work by van Dam \textit{et al}.  \cite{vandam03}. They also gave a reduction to the injective case based on a conjecture in   \cite{boneh95} that any string formed by $l$ subsequent values of $f$ is unique where $l>2\log^2 p$.  Gavinsky \textit{et al}. gave an efficient quantum algorithm in   \cite{gavinsky11} for solving the hidden shift problem for the average case Boolean functions $f:\mathbb{Z}_{2}^{n}\rightarrow \mathbb{Z}_{2}$. Ozols \textit{et al} gave another quantum algorithm for the Boolean hidden shift problem based on a quantum analogue of the rejection sampling defined in their paper  \cite{ozlos11}. Roetteler gave an efficient quantum algorithm in   \cite{roetteler10} for solving the hidden shift problem for several classes of the so-called bent functions. Later in   \cite{gavinsky11}, the hidden shift problem for all bent functions was solved as a special case of their algorithm. Bent functions are the Boolean functions $f(x):\mathbb{Z}_{2}^{n}\rightarrow \mathbb{Z}_{2}$ for which applying Hadamard transform to the function $f'(x):=(-1)^{f(x)}$ will yield Fourier coefficients of equal absolute value  \cite{rothaus76}. A complete characterization of bent functions seems to be a subtle task. However, it can be shown that bent functions do not exist for values of $n$ that are odd  \cite{rothaus76}. For large enough values of $n$ that are even, bent functions are guaranteed to exist and their count is at least $\Omega\left({2^{2^{n-1}+1/2{n \choose n/2}}}\right)$  \cite{carlet06}. 
\vspace*{1pt}
\subsection{Our results}
\setcounter{footnote}{0}\renewcommand{\thefootnote}{\alph{footnote}}
\noindent
 In the next section, we introduce a framework that we call \textit{injectivization}. We show that this tool can be used particularly for reducing the average case non-injective hidden shift problem for functions from any abelian or non-abelian group $G$ to any finite set to the injective hidden shift problem over the same group. Also, it can be used to reduce the (non-injective) hidden shift problem for bent functions to the injective hidden shift problem over the same group (which is $\mathbb{Z}_{2}^{n}$). We relate the applicability of this tool to a generalized notion of influence of the function.

These results about the hidden shift problem for the average case Boolean functions and bent functions and the relation to the function's influence simplify the main result in   \cite{gavinsky11}. We show that the Boolean hidden shift problem and the hidden shift problem for bent functions both reduce to Simon's problem since the injective hidden shift problem over $\mathbb{Z}_{2}^{n}$ admits a straightforward reduction to Simon's problem. Furthermore, these results answer an open question they ask, whether their methods can be generalized and adapted for the case of non-Boolean functions, as well. We do not use the methods in   \cite{gavinsky11}, but using our own method, we generalize the results in   \cite{gavinsky11} to functions whose range are arbitrary sets and are defined over groups of form $\mathbb{Z}_q^n$ with $q$ a constant prime power.
\vspace*{1pt}
\section{Injectivization}
\setcounter{footnote}{0}\renewcommand{\thefootnote}{\alph{footnote}}
\noindent
Injectivization is a process making it possible to transform two given non-injective functions defined over an arbitrary finite group into two injective functions defined over the same group while preserving the shift structure between them. The framework that we describe below is a way of constructing an injectivization process.

In this paper, we use $G$ to refer to an arbitrary finite group and $S$ to refer to an arbitrary finite set. We denote the $k$-th component of an $m$-tuple $V\in G^m$ with $v_k$.

The injectivization's input and output are specified in the following way:
\begin{itemize}
 \item Input: any function $f : G\rightarrow S$ and an $m$-tuple $V\in G^{m}$,
 \item Output: function ${f}_{V} : G\rightarrow S^m$ constructed in the following way:
\begin{equation}
{{f}_{V}}(x):=(f(xv_1),f(xv_2),\ldots,f(xv_m))\,. \label{eq2} 
\end{equation}
\end{itemize}
We say injectivization succeeds if ${f}_{V}$ is injective; otherwise it fails.
\vspace*{1pt}
\subsection{The average case non-injective hidden shift problem}
\setcounter{footnote}{0}\renewcommand{\thefootnote}{\alph{footnote}}
\noindent
To show that injectivization fails only with small probability when the input function $f:G\rightarrow S$ is chosen uniformly at random, in Theorem 1 we show that the probability of a collision (i.e., the existence of $x,y\in G$ such that ${{f}_{V}}(x)={f}_{V}(y)$) is small if $V$ has distinct components. Note that random variables ${{f}_{V}}(x)$ and ${{f}_{V}}(y)$ are not necessarily independent. We slightly abuse the definition of the non-injective hidden shift problem in Theorem 1 and Corollary 2. We make no promise that functions are not periodic.

\vspace*{12pt}
\noindent{\bf Theorem~1:} For arbitrary $V\in G^m$ with distinct components and for uniformly random function $f:G\rightarrow S$ the probability that ${{f}_{V}}$ is not injective is at most $\dfrac{\left| G \right|^2}{\left| S \right|^{\lceil{m/2}\rceil}}$.

\vspace*{12pt}
\noindent
{\bf Proof:} We will show that, for any distinct $x$ and $y$ in the domain, $\Pr[{{f}_{V}}(x) = {{f}_{V}}(y)] \le {1/\left| S \right|^{\lceil{m/2}\rceil}}$ and then the result follows from the union bound.

Let $x$ and $y$ be any two points in the domain of the function. If all of the components of $(xv_1,xv_2,\dots,xv_m)$ and $(yv_1,yv_2,\dots,yv_m)$ are distinct then it is clear that equality in each component is independent, so $\Pr[{{f}_{V}}(x) = {{f}_{V}}(y)] = {1/\left| S \right|^{m}}$.
However, the components need not all be distinct in which case there can be dependencies among components. To illustrate, consider the case where $G = \mathbb{Z}_{2}^{n}$, $|S|=2$ and $m=2$. We use additive notation temporarily. If $x = v_1$ and $y= v_2$ then $(x+v_1,x+v_2) = (0,v_1 \oplus v_2)$ and 
$(y+v_1,y+v_2) = (v_2\oplus v_1, 0)$, so a collision in the first component implies a collision in the second component.
Therefore, the probability of the collision ${{f}_{V}}(x) = {{f}_{V}}(y)$ is $1/2$ rather than $1/4$.

\begin{figure} [htbp]
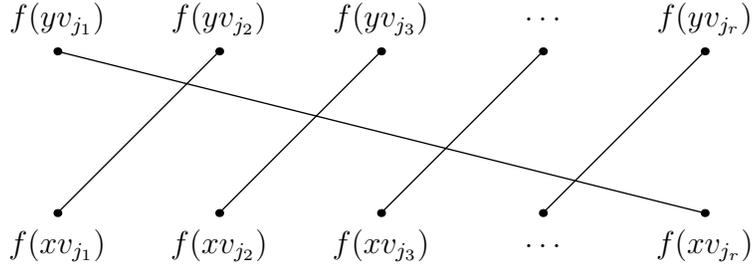

\vspace*{13pt}
\centering
\begin{asy}
size (10cm, 0);
draw ((0,0)--(100,100),black);
draw ((100,0)--(200,100),black);
draw ((200,0)--(300,100),black);
draw ((300,0)--(400,100),black);
draw ((0,100)--(400,0),black);

dot((0,100));
dot((0,0));
dot((100,100));
dot((100,0));
dot((200,100));
dot((200,0));
dot((300,100));
dot((300,0));
dot((400,100));
dot((400,0));

label ("$f(xv_{j_1})$", (0,-20));
label ("$f(yv_{j_1})$", (0,120));
label ("$f(xv_{j_2})$", (100,-20));
label ("$f(yv_{j_2})$", (100,120));
label ("$f(xv_{j_3})$", (200,-20));
label ("$f(yv_{j_3})$", (200,120));
label ("$\dots$", (300,-20));
label ("$\dots$", (300,120));
label ("$f(xv_{j_{r}})$", (400,-20));
label ("$f(yv_{j_{r}})$", (400,120));
\end{asy}
\vspace*{13pt}
\caption{\label{fig1}Components are shown with vertices and equal components are connected with an edge.}
\end{figure}

To address the general case, consider a maximal chain of dependencies:
\begin{align}\label{eq3}
\nonumber
xv_{j_1} & =  yv_{j_2} \\ 
\nonumber
xv_{j_2} & =  yv_{j_3} \\ 
\nonumber
&\setbox0\hbox{=}\mathrel{\makebox[\wd0]{\vdots}} \\
xv_{j_r} & =  yv_{j_{r+1}} \,. 
\end{align}
If $j_{r+1} = j_1$ then we have an $r$-cycle (the above example is a 2-cycle)(Fig. \ref{fig1}.).
Collisions in components $j_2,\dots,j_r$ of ${f}_V(x)$ and ${f}_V(y)$ occur independently; however if all these components collide, a collision in the component $j_1$ is implied (Fig. \ref{fig2}.).
Therefore, the probability of a collision among components $j_1, \dots, j_r$ is 
${1/\left| S \right|^{r-1}}$.
If, on the other hand, the chain is not cyclic then the probability of a collision among components $j_1, \dots, j_{r+1}$ is ${1/\left| S \right|^{r+1}}$.
Since all maximal chains of dependencies are disjoint, the probability of 
${{f}_{V}}(x) = {{f}_{V}}(y)$ is the highest when there are $m/2$ 2-cycles, when it is ${1/\left| S \right|^{\lceil{m/2}\rceil}}$ (Fig. \ref{fig3}.)$\blacksquare$\,.
%
\begin{figure} [htbp]
\centering
\begin{asy}
size (10cm, 0);
draw ((0,0)--(100,100),black);
draw ((100,0)--(200,100),black);
draw ((200,0)--(300,100),black);
draw ((300,0)--(400,100),black);
draw ((0,100)--(400,0),black);

draw ((100,0)--(100,100),dashed);
draw ((200,0)--(200,100),dashed);
draw ((300,0)--(300,100),dashed);
draw ((400,0)--(400,100),dashed);

dot((0,100));
dot((0,0));
dot((100,100));
dot((100,0));
dot((200,100));
dot((200,0));
dot((300,100));
dot((300,0));
dot((400,100));
dot((400,0));

label ("$f(xv_{j_1})$", (0,-20));
label ("$f(yv_{j_1})$", (0,120));
label ("$f(xv_{j_2})$", (100,-20));
label ("$f(yv_{j_2})$", (100,120));
label ("$f(xv_{j_3})$", (200,-20));
label ("$f(yv_{j_3})$", (200,120));
label ("$\dots$", (300,-20));
label ("$\dots$", (300,120));
label ("$f(xv_{j_{r}})$", (400,-20));
label ("$f(yv_{j_{r}})$", (400,120));
\end{asy}
\vspace*{13pt}
\caption{\label{fig2}Components are shown with vertices and equal components are connected with an edge. Dashed lines show the collisions.}
\end{figure}
\begin{figure} [htbp]
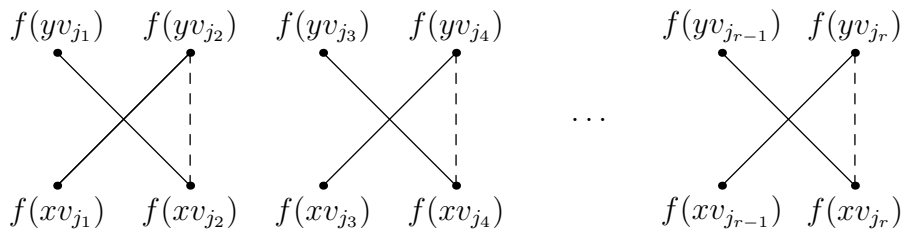

\centering
\begin{asy}
size (12cm, 0);
draw ((0,0)--(100,100),black);
draw ((100,0)--(0,100),black);
draw ((0,0)--(100,100),black);
draw ((200,0)--(300,100),black);
draw ((300,0)--(200,100),black);
draw ((500,0)--(600,100),black);
draw ((600,0)--(500,100),black);

draw ((100,0)--(100,100),dashed);
draw ((300,0)--(300,100),dashed);
draw ((600,0)--(600,100),dashed);

dot((0,100));
dot((0,0));
dot((100,100));
dot((100,0));
dot((200,100));
dot((200,0));
dot((300,100));
dot((300,0));
dot((500,100));
dot((500,0));
dot((600,100));
dot((600,0));

label ("$f(xv_{j_1})$", (0,-20));
label ("$f(yv_{j_1})$", (0,120));
label ("$f(xv_{j_2})$", (100,-20));
label ("$f(yv_{j_2})$", (100,120));
label ("$f(xv_{j_3})$", (200,-20));
label ("$f(yv_{j_3})$", (200,120));
label ("$f(xv_{j_4})$", (300,-20));
label ("$f(yv_{j_4})$", (300,120));
label ("$\dots$", (400,50));
label ("$f(xv_{j_{r-1}})$", (500,-20));
label ("$f(yv_{j_{r-1}})$", (500,120));
label ("$f(xv_{j_{r}})$", (600,-20));
label ("$f(yv_{j_{r}})$", (600,120));
\end{asy}
\vspace*{13pt}
\caption{\label{fig3}Components are shown with vertices and equal components are connected with an edge. Dashed lines show the collisions.}
\end{figure}

It is not hard to show that injectivization preserves the shift. More formally, pick an arbitrary $V\in G^m$ where $m$ is any positive integer. Pick functions $f,g:G\rightarrow S$. For any $s\in G$, it holds that $f( x )=g( sx )$ for all $x\in G$ if and only if ${{f}_V}( x )={{g}_V}( sx )$ for all $x\in G$. Furthermore, given oracles for $f,g$, it is straightforward to simulate a query to $f_V$ and $g_V$ efficiently, in both quantum and classical regime. Using these and Theorem 1, we obtain the following corollary.

\vspace*{12pt}
\noindent
{\bf Corollary~2:} Injectivization, when it succeeds, reduces an  instance of the non-injective hidden shift problem $f,g:G\rightarrow S$ to an instance of the hidden shift problem ${{f}_V},{{g}_V}:G\rightarrow S^{m}$ where $m$ is the number of $V$'s components. Injectivization fails with probability at most $\dfrac{\left| G \right|^2}{\left| S \right|^{\lceil{m/2}\rceil}}$ over the uniform random choice of $f$.

\vspace*{12pt}
\noindent
Theorem 1 specifies an upper bound on the failure rate of the injectivization process when the function $f$ is chosen uniformly at random. Injectivization process always fails when the input function is periodic since the output function also will be periodic. As a result, the failure rate when $f$ is a uniformly and randomly chosen function in Corollary 2 is also an upper bound on the failure rate when $f$ is a non-periodic uniformly and randomly chosen function. Using this and Corollary 2, the following corollary is trivial for polynomially large $m$:

\vspace*{12pt}
\noindent
{\bf Corollary~3:} Let $V\in G^m$ be composed of $m$ distinct components. Having $m\geq (4+\epsilon)\log_{\left| S \right|}{\left|G\right|}$ with an arbitrary constant $\epsilon>0$, an instance of the non-injective hidden shift problem $f,g:G\rightarrow S$ is reduced to an instance of the hidden shift problem ${{f}_V},{{g}_V}:G\rightarrow S^{m}$ with extremely high probability (asymptotically) over the uniform random choice of the non-periodic function $f$.

\vspace*{12pt}
\noindent
Theorem 2 in   \cite{gavinsky11} states that by the algorithms in   \cite{gavinsky11}, an average case exponential separation can be achieved. This result can be simplified by reducing the Boolean hidden shift problem to Simon's problem  \cite{simon97}:

\vspace*{12pt}
\noindent
{\bf Corollary~4:} The average case Boolean hidden shift problem reduces to Simon's problem using injectivization over $f,g:\mathbb{Z}_2^n \rightarrow \{0,1\}$ and then constructing the blackbox in Simon's problem $h:\mathbb{Z}_2^{n+1} \rightarrow \{0,1\}^m$ in the following way:
\begin{align}\label{eq4}
h( x_n x_{n-1}\ldots x_1 x_0)=\begin{cases} {{f}_V(  x_{n-1} x_{n-2}\ldots x_0 )}, & \mbox{if } x_n=0 \\ {{g}_V(  x_{n-1} x_{n-2}\ldots x_0 )}, & \mbox{if } x_n=1 \,.\end{cases} 
\end{align}

\vspace*{12pt}
\noindent
Gavinsky \textit{et al}. posed an open question in   \cite{gavinsky11} whether the methods they have used for solving the hidden shift over $\mathbb{Z}_{2}^{n}$ for the Boolean functions can be generalized and adapted for the case of non-Boolean functions. We have not used the method in   \cite{gavinsky11}, but we can say that using injectivization, as described above, we can reduce the average case non-injective hidden shift problem over $\mathbb{Z}_{2}^{n}$ to Simon's problem. Since in Simon's problem, it is not important for the functions to have range in binary strings, our functions need not be binary and they can have range in any finite set $S$. Furthermore, considering the domain to be the group $\mathbb{Z}_{q}^{n}$ with $q\geq 3$ a constant prime power, using injectivization, we can reduce the problem to the already solved injective case  \cite{friedl03,ivanyos08}.
\vspace*{1pt}
\subsection{Relation between the hidden shift problem and influence over the functions}
\setcounter{footnote}{0}\renewcommand{\thefootnote}{\alph{footnote}}
\noindent
We extend the notion of influence to the functions defined over any group $G$ and having range in any set $S$. The influence of $v$ over $f:G\rightarrow S$ is defined as $\gamma_v(f)=Pr_x[f(x)\ne f(xv)]$. When $G=\mathbb{Z}_{2}^{n}$ and $S=\{0,1\}$, this definition reduces to the conventional notion of influence. It is not hard to see that the function $f$ is periodic if and only if for some $v\in G\setminus \{1\}$: $\gamma_v=0$. Thus, the hidden shift problem with underlying functions $f,g$ is well-defined if the minimum influence of $f$, that is, $\gamma_{min}(f):=min_{v\in G\setminus\{1\}}(\gamma_v(f))$ is not zero.

\vspace*{12pt}
\noindent
{\bf Theorem~5:} For a uniformly at random chosen $V\in G^m$ and a function $f:G\rightarrow S$ the probability that ${{f}_{V}}$ is not injective is at most ${\dfrac{N}{2}\sum_{x\in G}}{(1-\gamma_{x})^m}\leq{N^2}{(1-\gamma_{min})^m}$.

\vspace*{12pt}
\noindent
{\bf Proof:} Let $N$ denote $|G|$. We define the matrix $A_{N\times N}$ according to
\begin{equation}\label{eq5}
A_{N\times N}=
\begin{bmatrix} 
f(x_0x_0)   & f(x_1x_0)   &\dots & f(x_{N-2}x_0) &f(x_{N-1}x_0)\\ 
f(x_0x_1)   & f(x_1x_1)   &\dots & f(x_{N-2}x_1) &f(x_{N-1}x_1)\\ 
\vdots 		   & \vdots 		   &\ddots & \vdots &\vdots\\ 
f(x_0x_{N-2})   & f(x_1x_{N-2})   &\dots & f(x_{N-2}x_{N-2}) &f(x_{N-1}x_{N-2})\\ 
f(x_0x_{N-1})   & f(x_1x_{N-1})   &\dots & f(x_{N-2}x_{N-1}) &f(x_{N-1}x_{N-1})
\end{bmatrix}
\end{equation}
where $x_0,x_1,x_2,\dots,x_{N-1}$ is an enumeration of elements  of $G$ in an arbitrary order.

For any two fixed and distinct rows $i,j$, the probability that their $k$-th element are equal is exactly $1-\gamma_{(x_i^{-1}x_j)}$ when $k$ is chosen uniformly at random. Thus, the probability that the strings of $m$ randomly chosen elements are equal is $(1-\gamma_{(x_i^{-1}x_j)})^m$ since the events are independent. Using union bound, it can be seen that the probability that any two strings of the form above are equal for any two distinct rows is at most 
\[{\sum_{i<j}}{(1-\gamma_{(x_i^{-1}x_j)})^m}={\dfrac{N}{2}\sum_{x\in G\setminus\{1\}}}{(1-\gamma_{x})^m}={\dfrac{N}{2}\sum_{x\in G}}{(1-\gamma_{x})^m}\]\[\leq{N^2}{(1-\gamma_{min})^m}.\] Based on the construction, this is an upper bound on the probability that ${f}_{V}$ is a non-injective function
$\blacksquare$\,.

\vspace*{12pt}
\noindent
In \cite{gavinsky11}, the number of queries needed by their algorithm to solve the hidden shift problem for functions of form $f,g:\mathbb{Z}_{2}^{n}\rightarrow \{0,1\}$ is shown to be related to the minimum influence of $f$. Interestingly, Theorem 5 relates the success probability of  injectivization to the same intrinsic feature of the function, that is, the minimum influence. To be precise, we are using the generalized notion of influence, but it remains the same for the case of binary functions. Hence, this gives an alternative proof that the average case Boolean functions can be injectivized when $V$ is chosen uniformly at random, due to a lower bound on the minimum influence of the majority of the Boolean functions in \cite{gavinsky11}. As a special case of this, using injectivization, it is possible to efficiently reduce the hidden shift problem for bent functions to the Simon's problem. Bent functions have a property called perfect nonlinearity, which means that, for any bent function $f:\mathbb{Z}_{2}^{n}\rightarrow \mathbb{Z}_{2}$ and for any non-zero $v\in \mathbb{Z}_{2}^{n}$, the function $f_v(x):=f(x)+f(x+v)$ is a balanced Boolean function  \cite{meier90}. This is equivalent to saying that $\gamma_v(f)=\gamma_{min}(f)=1/2$ for any non-zero $v$. Using Theorem 5 and a construction similar to Corollary 4, we have the following corollary:

\vspace*{12pt}
\noindent
{\bf Corollary~6:} Choosing $m>(2+\epsilon)n$ with an arbitrary constant $\epsilon>0$, using injectivization, the hidden shift problem for bent functions $f,g:\mathbb{Z}_{2}^{n}\rightarrow \mathbb{Z}_{2}$ reduces to the injective hidden shift problem $f_V,g_V:\mathbb{Z}_{2}^{n}\rightarrow \mathbb{Z}_{2}^m$ with high probability (asymptotically) which in turn reduces to Simon's problem. 
\vspace*{1pt}
\section{Classical complexity}
\setcounter{footnote}{0}\renewcommand{\thefootnote}{\alph{footnote}}
\noindent
We show that, the classical query complexity of the non-injective hidden shift problem when the underlying group is $\mathbb{Z}_{m}^{n}$ is high in the average case when $m$ is a constant number. For proving this bound, we benefit from some of the ideas in   \cite{gavinsky11}. 

First, we  define an artificial variant of the non-injective hidden shift problem which helps in proving the classical lower bound on the complexity of the average case non-injective hidden shift problem. We call this problem, the no-promise non-injective hidden shift problem. The only difference in this new problem is that we first pick $s\in G$ and oracle $g:G\rightarrow S$. Then the oracle $f:G\rightarrow S$ will be constructed according to (\ref{eq1}). The goal of the problem is to find $s$ given oracles $f$ and $g$.
In this problem, when $f$ and $g$ happen to be periodic functions, information theoretically it is not possible to choose the right $s$ with certainty among the many possible candidates. 

Similar to \cite{gavinsky11}, queries are made to the pair of functions $(f,g)$. This at most doubles the number of queries which is not important in the context of query complexity.

\vspace*{12pt}
\noindent
{\bf Theorem~7:} To solve a uniformly random instance of the no-promise non-injective hidden shift problem defined with a solution $s\in \mathbb{Z}_{q}^{n}$ and functions $f,g:\mathbb{Z}_{q}^{n}\rightarrow S$ with probability at least $1/2$, at least $\Omega\left( p_1^{n/2} \right)$ queries are needed when $q$ is a constant number and $p_1$ is the smallest prime divisor of $q$.

\vspace*{12pt}
\noindent
{\bf Proof:} Let $q=p_{1}^{k_{1}}\times p_{2}^{k_{2}}\dots \times p_{t}^{k_{t}}$ be the prime factorization of $q$ where $p_{1}<p_{2}<\dots<p_{t}$ holds. Let $T_1, T_2, \dots, T_m$ be an enumeration of all 1-dimensional subspaces of $\mathbb{Z}_{p_1}^{n}$. Since $p_1$ is a prime number, all subspaces have the same number of elements. Furthermore, the only common element between each two subspaces is $0$. These two imply $m=\dfrac{p_1^n-1}{p_1-1}$. 

We define the disjoint sets $S_{i}=T_{i}\setminus \{0\}$ for all $1\le i\le m$. We use the following notation: for $x\in \mathbb{Z}_{q}^{n}$, we define $x_{p_1}\in \mathbb{Z}_{p_1}^{n}$ such that $x_{p_1}=(x\text{ mod }p_1)$ where $\text{mod }p_1$ is carried out component-wise.

As a bonus to the classical computer, we provide a magical bell to it which rings if it makes the queries $X_1$ and $X_2$ and it happens that $(X_1-X_2)_{p_1}$ and $s_{p_1}$ are both in the same set $S_i$. If finding in which set $s_{p_1}$ lies, proves to be hard, then finding $s$ itself must be hard because of an obvious reduction from the latter to the former.

Without loss of generality, we assume $s_{p_1}\ne 0$. Let $Q_k=\{X_1,X_2,\dots,X_k\}$ be the places in which the queries are made after $k$ queries. Also, let $D$ be the set of all $i$'s  for which we know $s_{p_1}\notin S_i$ according to our queries and the magical bell. The sets $S_1, S_2, \dots, S_m$ are disjoint. This gives the important observation that knowing $D$ gives no information about the actual set to which $s_{p_1}$ belongs. More formally, we have 
\begin{equation}
\Pr\left[s_{p_1}\in S_{i}\vert i\notin D\right]=\frac{1}{m-|D|}\leq\frac{1}{\dfrac{p_1^{n}-1}{p_1-1}-k^{2}}.\label{eq6}
\end{equation}
Since conditioning on the queries does not provide any information, the best algorithm is to just randomly guess the set to which $s_{p_1}$ belongs. The best a classical computer can do is to eliminate $1+{k \choose 2}\leq k^2$ possible sets after $k$ queries. Hence, to be able to find the set to which $s_{p_1}$ belongs with probability at least $1/2$, it needs
to make at least $\Omega\left(p_1^{n/2}\right)$
queries $\blacksquare$\,.

\vspace*{12pt}
\noindent
{\bf Theorem~8:} To solve the non-injective hidden shift problem for functions $f,g:\mathbb{Z}_{q}^{n}\rightarrow S$ with probability at least $1/2+\epsilon$ classically, $\Omega\left( p_1^{n/2} \right)$ queries are needed in the average case, when $\epsilon>0$ is an arbitrary constant.

\vspace*{12pt}
\noindent
{\bf Proof:} The probability that $f$ is periodic is very small. More formally, for any fixed non-zero $ r\in \mathbb{Z}_{q}^{n}$, it holds that
\begin{equation}
\Pr[f(x+r)=f(x)\text{ for all }x] \leq \dfrac{1}{|S|^{{q^n}/2}}\label{eq7}
\end{equation}
where the probability of the event is the highest when the order of $r$ is 2. Hence, by the union bound, the probability of having a periodic function is at most  $\dfrac{q^n}{|S|^{{q^n}/2}}$ which is double exponentially small. This implies that the number of periodic functions is at most $R:=N\dfrac{q^n}{|S|^{{q^n}/2}}$ where $N:=|S|^{q^n}$ denotes the total number of functions.

As the name suggests, adding the promise that the functions are non-periodic makes the no-promise non-injective hidden shift problem the same as the injective hidden shift problem in definition. Since the number of periodic functions is negligible, the uniform probability distribution over the whole language, $U$, is extremely close in variation distance to the uniform probability distribution over non-periodic functions, $V$. More formally:
\begin{equation}
\|U-V\|=\dfrac{1}{2}\sum_{x\in L}|U(x)-V(x)|\leq \dfrac{1}{2}\left((N-R)(\dfrac{1}{N-R}-\dfrac{1}{N})+\dfrac{R}{N}\right)=\dfrac{R}{N}.\label{eq8}
\end{equation}
Using Theorem 7, it implies immediately that there should not exist a probabilistic classical Turing machine that makes less than $\Omega\left(p_1^{n/2}\right)$ queries and solves a uniformly random chosen instance of the non-injective hidden shift problem with probability at least $\dfrac{1}{2}+\dfrac{R}{N}$, otherwise we could use this Turing machine to violate Theorem 7 $\blacksquare$\,.
\vspace*{1pt}
\section{Conclusion}
\noindent
We developed a framework called injectivization which can be used for reducing some instances of the non-injective hidden shift problem over any group to the hidden shift problem for injective functions over the same group. In particular, we showed that this process succeeds, when we have an average case instance of the non-injective hidden shift problem and also when the underlying function is bent. We related the success probability of this process to a generalized notion of influence. In addition, we simplified the main result of   \cite{gavinsky11} and also used this framework to address an open question of   \cite{gavinsky11} by generalizing their results to the hidden shift problem for functions $f,g:\mathbb{Z}_q^n \rightarrow S$ where $S$ is an arbitrary set and $q$ is a constant prime power. We also proved that the average case classical complexity of this problem for any constant $q$ is high.

\vspace*{1pt}
\section*{Acknowledgements}
\noindent
We are grateful to Richard Cleve for helpful discussions and his revision of this paper and to Andrew Childs for fruitful discussions. We also thank Dmitry Gavinsky for helpful email correspondence and the anonymous referee for the comments that significantly improved this work.

\end{document}